Simple Explanation of Fermi Arcs in Cuprate Pseudogaps: A Motional Narrowing Phenomenon

**ABSTRACT: ARPES measurements on underdoped cuprates above the superconducting transition temperature exhibit the unique phenomenon of Fermi arcs, gapless arcs of Fermi surface around the nodal points of the superconducting gap, which terminate before reaching the antinodes. We show that this phenomenon is easily explained (including its temperature dependence and observed hole-electron asymmetry) as the natural consequence of a time-fluctuating d-wave gap.**

ARPES measurements in the vortex liquid[1] part of the pseudogap region of underdoped BISSCO cuprates show that the spectrum retains an energy gap of d symmetry, but that around the nodal points that gap appears to have collapsed, leaving a finite arc of apparently true Fermi surface, which simply terminates. In the antinodal region the gap remains nearly as large as in the superconductor.[2,3] In the experiments there is no indication that this arc represents a part of a true Fermi surface pocket, but this has not prevented the publication of various theoretical interpretations in such terms.[4,5]

Whatever other properties this region of the pseudogap phase may have, the one thing we know with certainty is that it is a phase fluctuating d-wave superconductor, since it is separated from the superconductor by an x-y symmetry phase transition and no other visible phase transition intervenes.[6] This has been amply confirmed in a sequence of experiments on the Nernst effect and nonlinear diamagnetism by Ong's group,[7] and discussed theoretically by the present author.[8] From a crude theory of the gap fluctuations, we have estimated that outside of the immediate vicinity of Tc their correlation time is of order $(kT/h)^{-1}$.[9]

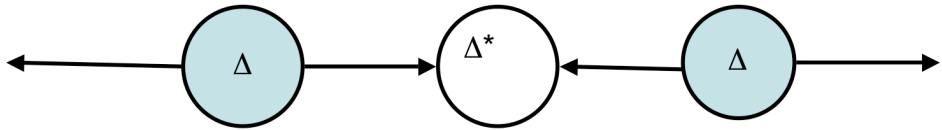

**quasiparticle:electron->hole->electron etc**

Figure 1: Near the Fermi surface, the quasiparticle is a resonant splitting of hole and electron eigenstates

Near the Fermi momentum, the BCS gap can be thought of as a splitting of the energy levels of the electron by the anomalous self-energy $\Delta_k$, which causes the quasiparticle to resonate between electron and hole states of the same momentum. (See Fig 1). The behavior of the quasiparticles may be described by a simple effective single-particle Hamiltonian,

$$H_k = \begin{pmatrix} \varepsilon_k & \Delta_k \\ \Delta^*_k & -\varepsilon_k \end{pmatrix} \quad [1]$$

for each momentum near the Fermi surface, acting on the two states $c^*_k$ and $c_{-k}$. $\varepsilon$ is the energy relative to the Fermi energy $\mu$.

The ARPES spectrum is proportional to the Fourier transform of the Green's function $\quad G_h(k,t) = \langle c_k^*(t) c_k(0) \rangle, \quad$ [2] where the average is taken in the thermal equilibrium state, and t>0: the photon can only create holes. Thus we need to know the spectrum of the single quasiparticle in the presence of a time-varying anomalous self-energy: the gap parameter $\Delta_k$ is varying in

time. As a result, the Hamiltonian [1] becomes time-dependent. Effectively, the Hamiltonian becomes [1] with
$\Delta = \Delta_0 e^{i\theta(t)}$ with $\Delta_0$ real, and $\Delta^* = \Delta_0 e^{-i\theta(t)}$. [3]
θ is a random function of time, and in most of the pseudogap region (not close to Tc) its correlation time is close to h/kT, as remarked in ref 8:

$$\left\langle e^{i(\theta(t+\tau)-\theta(t))} \right\rangle \cong e^{-\Gamma\tau} \quad where \quad \Gamma \approx kT/\hbar \quad [4]$$

In [4] we are assuming the statistics of the gap variations to be Markovian, which we do for mathematical convenience; the motion of the phase is caused by the random tangle of vortices rotating around each other and is more correlated than [4] at short times, less so at long—perhaps better modeled by a Gaussian than a Lorentzian phase modulation—but this is known to have little effect on the narrowing phenomenon. [10]

Now
$$c^*_k(t) = U^{-1}(t) c^*_k(0) U(t) \quad [5] \text{ and the time-development operator}$$
U is defined by $i\hbar \partial U/\partial t = HU$, which is solved by

$$U(t) = T[e^{i\int_0^t H dt}] \quad [6]$$

Here T is the time-ordering operator which is necessary if H(t) at different t do not commute. But for the important case here, which is for states right at the Fermi level ($\varepsilon_k=0$), H(t) is always of the same form and commutes. In fact, the semiclassical approximation of dropping the time-ordering operator T seems always (ref 10) to give good results in line-narrowing problems, and will in this one.

Reference [10] provides a "template" for the present problem, in that one of the cases on which explicit results are known is the narrowing out of the splitting of a doublet such as the ammonia inversion line. But the analogy is not perfect: in the classic cases that paper was written for, kT is large compared to the splitting and the motion which causes the narrowing is simply random fluctuations of a "bath", which could be modeled as random flipping of the frequency from one member of the doublet to the other. In the present case, the narrowing transition takes place when the doublet splitting—the gap—is of order kT so the energy dependence of the flipping process cannot be neglected. In fact, as a consequence of detailed balance we reveal an asymmetry in the spectrum which is a characteristic feature of the observations[11].

FORMALISM

First let us adapt the formalism of ref 10 to the present problem, which should give the correct answer for the nodal region where $\Delta$<kT, that is, in the Fermi arc. We confine our interest to the quasiparticles right at the Fermi surface, that is with $\varepsilon$=0, because these are the quasiparticles which define the gap. Here we can model the stochastic effect of random phase fluctuations of $\Delta$ as random flips between the two possible quasiparticle states, with energies $\pm\Delta$. U(t) evolves during the intervals with the corresponding frequency $\pm\omega_0 = \pm\Delta/h$.

First let us present the appropriate formalism from ref 10. A system with two frequencies, $\pm\omega_0$, is assumed to randomly flip back and forth between the two at a rate $\Gamma$. This process can be described by a 2x2 matrix for the probability that the frequency is $\omega$ at the beginning of a short interval dt and $\omega$' at the end of it:

$$W_{\omega\omega'}(dt) = \begin{pmatrix} 1-\Gamma dt & \Gamma dt \\ \Gamma dt & 1-\Gamma dt \end{pmatrix} \quad (\omega,\omega'=\pm\omega_0)$$

$$= 1 + \Pi dt$$

[7]

During the interval dt U(t) changes by

$$dU = \begin{pmatrix} i\omega_0 & 0 \\ 0 & -i\omega_0 \end{pmatrix} dt, \qquad [8]$$

so combining [7] and [8] into a differential equation for the average U(t), we obtain

$$\langle U(t) \rangle = \exp\begin{pmatrix} i\omega_0 - \Gamma & \Gamma \\ \Gamma & -i\omega_0 - \Gamma \end{pmatrix} t \qquad [9]$$

Inserting this into [2] and using [5], we see that the Green's function is the sum of two simple exponentials in t,
$G = ae^{\lambda_1 t} + be^{\lambda_2 t}$, with $\lambda_{1,2}$ being the eigenvalues of the matrix in [9]
The spectrum then, is the sum of two lorentzians. The eigenvalues are

$$\lambda = -\Gamma \pm \sqrt{\Gamma^2 - \omega_0^2} \qquad [10]$$

When $\Gamma$, the flip rate, $\gg \omega_0$, we have the usual narrowing phenomenon: the smaller of these two eigenvalues, $\approx -\omega_0^2/2\Gamma$ represents a much slower-decaying exponential, i e a sharp line at $\omega = 0$.
This line turns out to contain most of the amplitude.

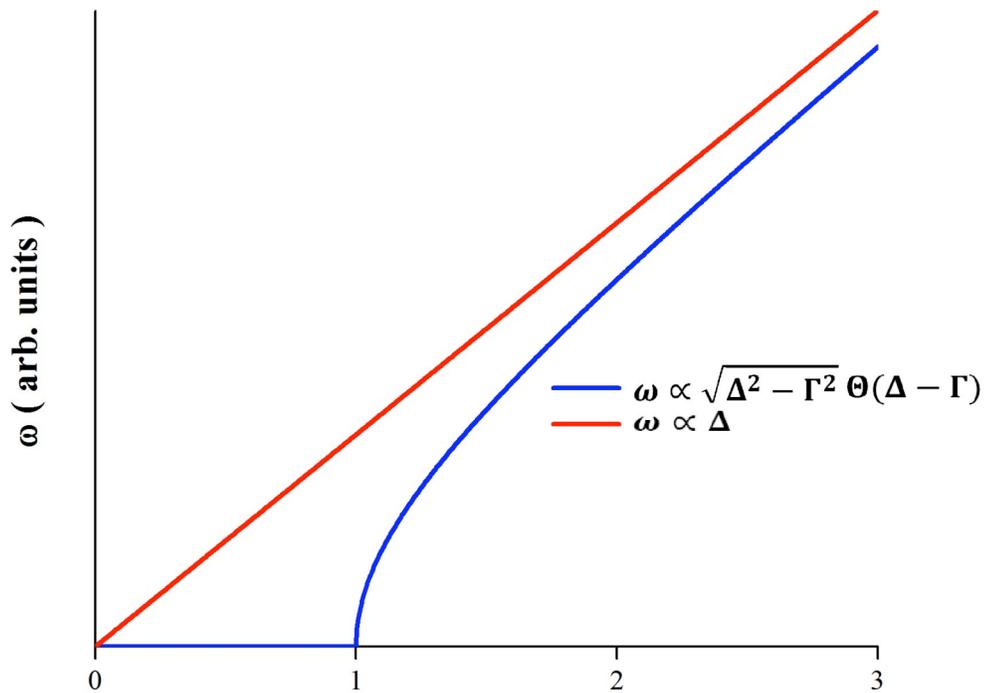

**Figure 2** Imaginary part of eigenvalue = line center frequency

When $\Gamma \ll \omega_0$, there are two equal lines at $\pm\omega_0$ decaying at the rate $\Gamma$. In the intermediate case, there is a rather sharp transition between the two cases (see Fig 2) and to a rough approximation, this transition can be thought of as the end of the "Fermi arc".

The actual behavior in our case, where the transition takes place when the energies are comparable to kT, is somewhat more complicated. We must take into account that the hopping process between the two states is also responsible for establishing thermal equilibrium: a downward jump in energy is more probable than an upward one, and the populations of the two states differ. (see Fig 3) . It is essential to our thinking to realize that any given quasiparticle scatters back and forth between the two states and so has no permanent identity as above or below the gap.

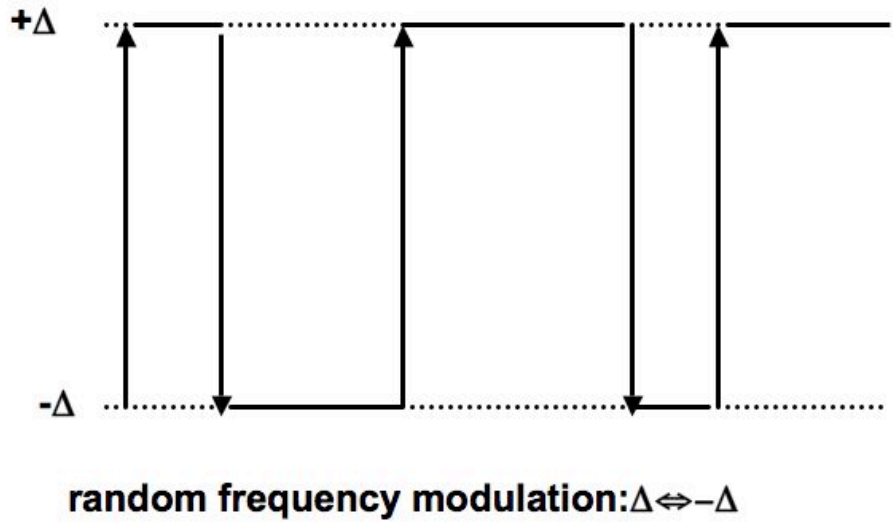

random frequency modulation: $\Delta \Leftrightarrow -\Delta$

**figure 3a**

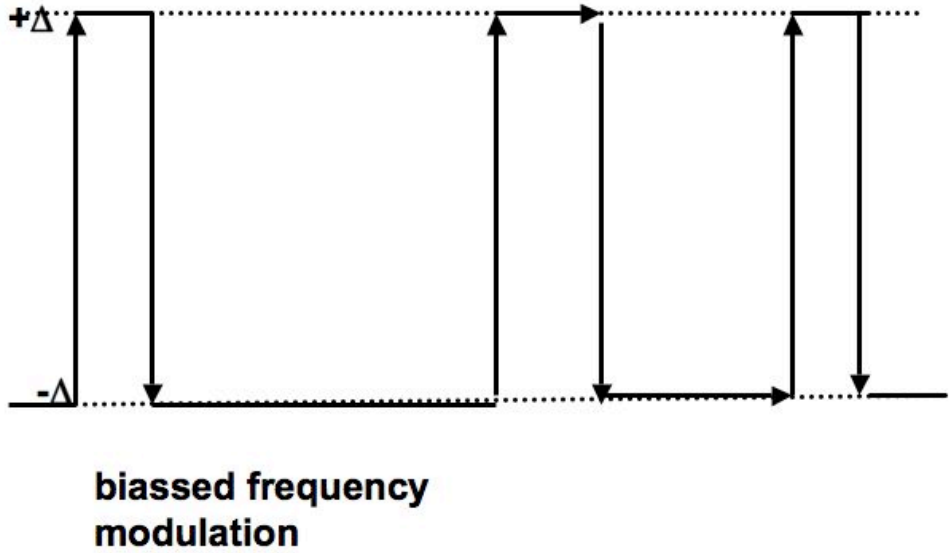

biassed frequency modulation

**figure 3b**

The probability matrix $\Pi$ for the scattering is not symmetric, since it must lead to the thermal distribution

$P(\pm\omega_0) \cong 1/(1+e^{\pm\beta\omega_0})$ where $\beta = 1/kT$    [11]

To get this distribution, $\Pi$ must be

$$\Pi = \begin{pmatrix} -\Gamma & \Gamma' \\ \Gamma & -\Gamma' \end{pmatrix} \text{ with } \frac{\Gamma}{\Gamma'} = e^{\beta\omega_0} \quad [12]$$

The eigenvalues are not as simple as [10]:

$$\lambda = -(\Gamma+\Gamma')/2 \pm \sqrt{(\Gamma+\Gamma')^2/4 - \omega_0^2 + i\omega_0(\Gamma-\Gamma')} \quad [13]$$

This formula satisfies all the limiting cases; for instance, when both Γ's are large, it gives us a narrowed line at the average frequency. When $\omega_0$ is large, it gives two lines, but with the positive frequency one having the greater breadth by a factor $e^{\beta\omega}$. The behavior near the transition is complex, but clearly asymmetrical between the two lines; what one must keep in mind is that the eigenvalue for the narrow, strong line evolves continuously to that for the low-energy one, and the broad, weak line becomes the high-energy one, and is still broader even when it is well past the crossover region. This behavior agrees completely with Peter Johnson's observations on the lack of symmetry between the two regions of the spectrum. The amplitude ratio can be compensated by multiplying up by the Fermi factor, but the asymmetry in breadth is over and above that.

The behavior is complicated but as an illustrative example I calculated out the transitional case

$$\omega_0 (=\Delta) = (\Gamma + \Gamma')/2 = \beta^{-1}$$

for this case the imaginary parts of the two eigenvalues

(the energies) are equal and opposite at about

$\pm 0.48\omega_0$, while the widths differ by a factor about 3.

The width of the higher-energy eigenvalue is three times its energy (about 1.5 $\omega_0$) and it would be hard to identify the central energy as shifted from zero, while for the lower-energy eigenvalue the width and shift are comparable.

CONCLUSIONS

It seems that the mysterious "Fermi arc" phenomenon is easily explained as a natural consequence of the nature of the pseudogap-vortex liquid state. Gaps around the nodal points are too small to survive the rapid phase fluctuations of the anomalous self-energy, but the large gaps at the antinodes remain and control the symmetry of the gap structure. A natural explanation appears of the observation in ref. 3 that the edge of the arc occurs when $\Delta \sim kT$. This narrowing effect is a nice confirmation of the nature of the pseudogap state, a confirmation which is strengthened by the observation in ref 11 of strong asymmetry of the breadths of states above and below the gap.

I would like to acknowledge discussions of the Fermi Arc data with Mike Norman, J-C Campuzano, and Peter Johnson, Peter's sharing of prepublication data, and discussions of the pseudogap state with A Yazdani and N P Ong.

[1] P W Anderson, Phys Rev Lett 100, 215301(2008)
[2] M Norman, H Ding, M Randeria, J-C Campuzano, et al, Nature 392, 157 (1998)
[3] A Kanigel et al, Nature Physics 2, 447 (2007)
[4] R M Kokik, T M Rice, A Tsvelik, cond-mat/0511268
[5] T D Stanescu, G Kotliar, cond-mat/0508302
[6] M J Salamon, et al, Phys Rev B47, 5520 (1993)
[7] N P Ong, and Yayu Wang, Physica 408-410C, 11 (2004); Y Wang, Lu Li, et al, Phys Rev Lett 97, 247002 (2005)
[8] P W Anderson, ref [1]
[9] P W Anderson, cond-mat/0603726
[10] PWA, J Phys Soc Jap 9, 316 (1954)
[11] H B Yang, P D Johnson, et al, submitted to Nature (2008)